\documentclass[preprint]{emulateapj}

\slugcomment{Accepted for publication in the Astrophysical Journal}
\shortauthors{Zaritsky et al.}
\shorttitle{}

\begin{document}
\title{Evidence for Two Distinct Stellar Initial Mass Functions}
  
\author{Dennis Zaritsky}
\affil{Steward Observatory, University of Arizona, 933 North Cherry Avenue, Tucson, AZ 85721}

\author{Janet E. Colucci}
\affil{Department of Astronomy and Astrophysics, 1156 High Street, UCO/Lick Observatory, University of California, Santa Cruz, CA 95064}

\author{Peter M. Pessev}
\affil{Gemini South Observatory, c/o AURA Inc., Casilla 603, La Serena, Chile}

\author{Rebecca A. Bernstein}
\affil{Department of Astronomy and Astrophysics, 1156 High Street, UCO/Lick Observatory, University of California, Santa Cruz, CA 95064}

\author{Rupali Chandar}
\affil{Department of Physics and Astronomy, The University of Toledo, 2801 West Bancroft Street, Toledo, OH, 43606}

\email{dennis.zaritsky@gmail.com}

\begin{abstract} 
We present velocity dispersion measurements of 20 Local Group stellar clusters ($7 < $ log(age [yrs]) $< 10.2$) from integrated light spectra and examine the evolution of the stellar mass-to-light ratio, $\Upsilon_*$.
We find that the clusters deviate from the evolutionary tracks corresponding to simple stellar populations drawn from standard stellar initial mass functions (IMFs). The nature of this failure, in which $\Upsilon_*$ is at first underestimated and then overestimated with age,  invalidates potential simple solutions involving a rescaling of either the measured masses or modeled luminosities. A range of possible shortcomings in the straightforward interpretation of the data, including subtleties arising from cluster dynamical evolution on the present day stellar mass functions and from stellar binarity on the measured velocity dispersions, do not materially affect this conclusion given the current understanding of those effects. Independent of further conjectures regarding the origin of this problem, this result highlights a basic failing of our understanding of the integrated stellar populations of these systems. We propose the existence of two distinct initial mass functions, one primarily, but not exclusively, valid for older, metal poor clusters and the other for primarily, but not exclusively, younger, metal rich clusters. The young (log(age [yrs])$<$9.5) clusters are well-described by a bottom-heavy IMF, such as a Salpeter IMF, while the older clusters are better described by a top-heavy IMF, such as a light-weighted Kroupa IMF, although neither of these specific forms is a unique solution. The sample is small, with the findings currently depending on the results for four key clusters, but doubling the sample is within reach.
\end{abstract}


\section{Introduction}
\label{sec:intro}

The stellar mass of a galaxy is intricately connected to that galaxy's environment \citep{kauffmann04}, metal abundance \citep{tremonti}, star formation history \citep{bundy}, dark matter halo mass \citep{vdb},  and just about any other principal characteristic one cares to consider. Yet, we rely on estimates of stellar masses that are potentially rife with systematic errors. Those estimates hinge on our poor knowledge of a galaxy's star formation history and are further predicated on two aspects of stellar evolution that are poorly understood: stars' behavior during the phase(s) of their life at which they are at their most luminous and 
the initial distribution of stellar masses (the initial mass function or IMF). Despite these various complications, we utilize simple models to estimate stellar masses because dynamical measurements are technically difficult to obtain and complicated by the presence of dark matter. Even when a dynamical mass estimate is available, modeling is required to determine the relative contributions of stars and dark matter to the total mass.  Given that any apparent failure of such models impugns our understanding of either stellar evolution or the IMF, and therefore affects many aspects of our study of the extragalactic universe, testing these models on the simplest possible stellar systems is paramount.

Discrepancies between models and observations, when found, are commonly attributed to deviations in the IMF from the adopted prescription \citep[for some recent examples see][]{vandokkum08,dabringhausen09,treu10,vandokkum,dutton} rather than to either the star formation history or stellar evolutionary models.
Direct measurements of the initial mass function are difficult for various reasons \citep[see][for a review]{bastian}, particularly over the full range of environments and conditions. Various ``standard" descriptions of the IMF are often used to span the range of possibility although given the simple mathematical descriptions used for the IMF it is not evident that such an approach is complete. Some of the IMFs in widest use include the original power-law description by \cite{salpeter}, which is generally acknowledged to be a poor fit to current data for nearby regions but remains in use as a benchmark \citep{bastian}, the multisegmented power-law \citep{kroupa}, and forms closer to log normal \citep{miller, chabrier}. It is often quite difficult to distinguish between multisegmented and log-normal variants \citep{dab}. Among the most direct measurements of the IMF, the evidence currently points to a universal IMF that is a power law with index close to  Salpeter's original value for stellar masses above a few solar masses and log normal or shallower for lower mass stars \citep{bastian}, despite hints of variations in other, more extreme, environments or epochs that are observationally less accessible \citep[for examples see][]{rieke, dave, desika}. In summary, the range of standard IMFs,  let alone the potential for variations across different environments, results in stellar mass uncertainties of at least a factor of several, which is unsatisfactory when testing models whose purpose is to measure differences in galaxy stellar masses to comparable or better precision.

To avoid the problems associated with testing these models against the observed properties of galaxies, with their complex stellar populations and ever present dark matter, we will test whether the models can reproduce the properties of local stellar clusters. This approach is not new. The challenge in implementing such a test is finding populations of clusters for which one can dynamically measure the cluster masses and also sample the relevant parameter ranges, principally cluster age, but also metallicity. To date, $\sim 90$\% of the local clusters with kinematic measurements, from which masses can be derived, are old, log(age [yrs]) $> 10$ \citep{mclaughlin, kruijssen09}.
The remainder tend to be extremely young and therefore subject to questions about whether they are dynamically relaxed \citep{goodwin}. Because systematic effects can bias the estimated masses either high or low as a group, a population of clusters all at one age does not provide a strong test of the models. To test the evolutionary predictions of models, including the effects of dynamical evolution, investigators have turned to the populations of clusters in galaxies outside the Local Group \citep{rejkuba, kruijssen08}. However, distance quickly diminishes our ability to measure the internal properties of clusters, particularly age. Here, we seek, using Local Group clusters, to measure the evolution of the stellar mass-to-light ratio, $\Upsilon_*$, for a set of clusters with well measured structural properties, ages, and chemical abundances, and to determine if  models of a simple stellar population with a universal IMF can produce a match to that evolution.

A large compilation of  Local Group stellar cluster data was published by \cite{mclaughlin}. The particular focus of that study was on the radial surface brightness profiles of the clusters and emphasis was placed on homogenizing the data obtained from various sources, thereby making it useful for subsequent studies as well. In addition, when available, they collated velocity dispersions and produced dynamical models. However, as we alluded to before, out of the 153 clusters presented in that study, measurements of the internal kinematics exist for only 57, and, of those, only 6 are younger than 10 Gyrs old. They compared their dynamically measured $\Upsilon_*$ to those derived from stellar population models, using both a variety of IMFs and two different evolution codes \citep{pegase, bc}. Although the mean ratio of $\Upsilon_*$ from their dynamical models to that from their preferred
stellar population models is $0.82\pm0.07$ (formally 2.5$\sigma$ discrepant with a ratio of 1), they do not stress this as a significant disagreement, presumably because they appreciate that systematic errors in their mass scale could be $\sim$20\%. Subsequently, \cite{kruijssen09} alleviated the apparent discrepancy by showing that due to the internal dynamical evolution of the clusters, which causes the preferential loss of low mass stars,  the measured values of $\Upsilon_*$ are expected to be lower than those calculated from stellar population models. Nevertheless, even if the models, once one includes dynamical evolution, are in quantitative agreement for this set of clusters, this result provides support for the models only at a single age. 

To expand the age range of the sample of clusters with dynamically measured values of $\Upsilon_*$ and define a homogenous sample for study, we undertook to measure the velocity dispersions of 22 Local Group stellar clusters. Here we present our measurements, using integrated spectra, of the internal velocity dispersions of 20 of those clusters that span $7 < \log( {\rm age}) < 10.2$ (for two we failed to obtain a measurement). Of those, we use the 18 
that are also in the compilation of \cite{mclaughlin} for 
our subsequent analysis. In \S2 we discuss the required data, both that drawn from \cite{mclaughlin} and our own observations. We present the observational details, the data reduction, and our measurements of the internal velocity dispersions. In \S3 we then convert these velocity dispersion measurements to mass estimates. In \S4, we present and discuss the behavior of $\Upsilon_*$ with age, and evaluate the possible effects of dynamical evolution and binary stars on these results. In \S5 we focus on our preferred hypothesis for the origin of the failure of the simple stellar population models to describe the behavior of $\Upsilon_*$ with age --- the existence of at least two distinct IMFs.  We summarize this study in \S7. 
   
\section{The Data}

Our spectroscopic data come from a set of observations taken with the Las Campanas du Pont telescope (100-inch) and the Magellan Clay telescope (6.5m). All of the du Pont data and some of the Magellan data were obtained prior to the formulation of the current work and were meant primarily for an investigation of the chemical abundance patterns within clusters \citep{colucci1,colucci2}. For the bulk of the Magellan observations that are the core of this work, 
we selected clusters from the compilation provided by \cite{mclaughlin}. The one exception is NGC 1718, which was observed (Nov. 2006) as part of the earlier work at Magellan. We use the compilation to select a range of clusters with the necessary ancillary data (age, half light radius, luminosity, modeled $\Upsilon_*$ from stellar population models, and metallicity). From that list, we selected the clusters in Table \ref{tab:clusters} that were observed with Magellan on the basis of their surface brightness within the half-light radius (which enables us to obtain high S/N, high dispersion spectra), with the additional requirement that we span the range of ages. Of the clusters that satisfy these criteria best, 6 already had measurements of the line-of-sight velocity dispersion, $\sigma$, contained in the published compilation. Nevertheless, we observed those systems as well to obtain a homogenous, and homogeneously analyzed, set of velocity dispersion measurements. Finally, we also use the set of old clusters (age $>$ 10 Gyr) with velocity dispersion measurements from the published compilation when we examine the question of dynamical evolution and its effect on $\Upsilon_*$ and discuss the possible origin of distinct IMFs. When model data are used from the compilation, we choose results obtained using the Wilson models \citep{wilson}, which \cite{mclaughlin} demonstrate are superior in fitting the radial surface brightness profiles of these clusters. Nevertheless, we tested whether our measurement of $\Upsilon_*$ were affected by 
this choice and the results are discussed in \S3.

\subsection{Ages, Photometry, and Stellar Population Models}

We adopt the ages, photometry, and stellar population results compiled by \cite{mclaughlin} without modification. The ages are derived from analyses of color-magnitude diagrams, compiled but not analyzed by those authors. The uncertainties are therefore somewhat heterogeneous and, as always, dependent on the age of the cluster. We will broadly adopt a 20\% uncertainty in the age in our Figures. The photometric measures, half-light radius, $r_h$, and mean surface brightness within the half-light radius, $I_h$, were calculated and tabulated by \cite{mclaughlin} from their surface brightness fits and uncertainties are provided. The uncertainties in our dynamical estimates of $\Upsilon_*$ are dominated by our uncertainties in the velocity dispersion $\sigma$, partly because the fractional errors are larger on $\sigma$ than on any other of the required parameters and partly because the mass depends on $\sigma^2$.  Finally, \cite{mclaughlin} present the results from a set of stellar population models. From the set they provide, we select those obtained with the \cite{bc} models and either a standard Salpeter IMF \citep{salpeter} or the often-used Chabrier disk IMF \citep{chabrier} and the \cite{pegase} (PEGASE) models and a \cite{kroupa} IMF. We adopt the uncertainties quoted in the compilation, although stochastic effects (how well sparsely populated features in the CMD are sampled) can lead to significant deviations from the expected photometric properties. For our clusters, which have $3.5 \times 10^4 < M < 5.0 \times 10^5 M_\odot$, as we will show below, the role of stochasticity in colors is potentially significant, although it is dramatically larger for $M < 10^4 M_\odot$ \citep{massclean}, and some scatter in the total luminosity is also expected.

We will also go beyond the model results provided by \cite{mclaughlin} by utilizing the models provided by \cite{anders}, which address the issue of the dynamical evolution of stellar clusters. We discuss these in detail in \S \ref{sec:dynamical}.

\begin{deluxetable*}{lllrrrrrrrrrr}
\tablewidth{0pt}
\tablecaption{Stellar Cluster Data}
\tablehead{
\colhead{NGC} &
\colhead{Host} &
\colhead{Tel.} &
\colhead{$t_{exp}$} &
\colhead{log(age)} &
\colhead{log(L$_{V}$)}&
\colhead{$\langle Fe/H \rangle$} &
\colhead{$r_h$} &
\colhead{log(I$_h$)} &
\colhead{$\sigma$} &
\colhead{$\Upsilon_{*}$} &
\colhead{$\Upsilon_{*,S}$} & 
\colhead{$\Upsilon_{*,C}$} \\
&&&[s]&[Gyr]&[L$_{\odot}$]&&[pc]&$[L_\odot pc^{-2}]$&km s$^{-1}$&$[\odot]$&$[\odot]$&$[\odot]$\\
}
\startdata
0121 & SMC &  Clay(10/11) &  9000 & 10.08 & 5.34 &$-$1.71 &  5.7 & 3.04 & $4.16^{+0.21}_{-0.21}$&  0.50$^{+0.05}_{-0.05}$ & 2.98 & 1.77 \\
0411 & SMC &  Clay(10/11) & 10800 &  9.15 & 4.90 &$-$0.68 &  7.1 & 2.39 & $3.29^{+0.15}_{-0.15}$&  1.10$^{+0.10}_{-0.10}$ & 0.73 & 0.41 \\
0416 & SMC &  Clay(10/11) &  5400 &  9.84 & 5.15 &$-$1.44 &  4.5 & 3.03 & $3.70^{+0.26}_{-0.27}$&  0.49$^{+0.49}_{-0.49}$ & 2.10 & 1.18 \\
0458 & SMC &  Clay(10/11) & 10800 &  8.30 & 4.98 &$-$0.23 &  6.0 & 2.63 &                    ...&                     ... & 0.24 & 0.14 \\
1711 & LMC &  C100 &  18350     &  7.70 & 5.49 &$-$0.57 &  5.5 & 3.21 & $5.55^{+1.09}_{-1.16}$&  0.61$^{+0.26}_{-0.23}$ & 0.10 & 0.06 \\
1718 & LMC &  Clay &  19560     &  9.30 & 4.80 &$-$0.42 &  6.4 & 2.38 & $4.93^{+0.22}_{-0.23}$&  2.80$^{+0.26}_{-0.26}$ & 1.19 & 0.66 \\
1831 & LMC &  Clay(2/11)& 10900 &  8.20 & 5.30 &   0.01 &  7.7 & 2.72 & $2.97^{+0.28}_{-0.32}$&  0.39$^{+0.08}_{-0.08}$ & 0.32 & 0.19 \\
1856 & LMC &  Clay(2 \& 10/11) &  3600 &  8.12 & 6.10 &$-$0.52 & 18.7 & 2.75 & $4.85^{+0.48}_{-0.53}$&  0.40$^{+0.08}_{-0.08}$ & 0.16 & 0.10 \\
1860 & LMC &  Clay(10/11) &  6300 &  8.28 & 5.49 &$-$0.52 & 35.9 & 1.59 &                    ...&                     ... & 0.21 & 0.12 \\
1866 & LMC &  C100 &   41230    &  8.12 & 5.93 &$-$0.50 &  9.9 & 3.14 & $6.50^{+0.44}_{-0.53}$&  0.55$^{+0.08}_{-0.09}$ & 0.16 & 0.10 \\
1868 & LMC &  Clay(2 \& 10/11) &  7200 &  8.74 & 4.97 &$-$0.50 &  3.2 & 3.16 & $4.05^{+0.17}_{-0.18}$&  0.62$^{+0.05}_{-0.05}$ & 0.41 & 0.23 \\
1916 & LMC &  C100 &  22785     & 10.11 & 5.51 &$-$2.08 &  2.0 & 4.10 & $9.01^{+0.27}_{-0.30}$&  0.57$^{+0.03}_{-0.03}$ & 3.14 & 1.90 \\
1978 & LMC &  C100 &  50150     &   ... &  ... &    ... &  ... &  ... & $5.48^{+0.22}_{-0.22}$&                     ... &  ... & ...  \\
2002 & LMC &  C100 &   18357    &   ... &  ... &    ... &  ... &  ... & $9.22^{+0.33}_{-0.36}$&                     ... &  ... & ...  \\
2005 & LMC &  C100 &  22010     & 10.11 & 5.06 &$-$1.92 &  2.0 & 3.68 & $7.46^{+0.40}_{-0.48}$&  1.05$^{+0.12}_{-0.13}$ & 3.12 & 1.88 \\
2019 & LMC &  C100 &  20218    & 10.11 & 5.26 &$-$1.81 &  2.6 & 3.63 & $7.29^{+0.30}_{-0.31}$&  0.85$^{+0.07}_{-0.07}$ & 3.11 & 1.87 \\
2031 & LMC &  Clay(2/11) & 10800 &  8.20 & 5.53 &$-$0.52 & 10.9 & 2.66 & $3.76^{+0.39}_{-0.43}$&  0.51$^{+0.11}_{-0.11}$ & 0.18 & 0.11 \\
2100 & LMC &  C100 &   20800    &  7.20 & 5.91 &$-$0.32 &  5.0 & 3.72 & $8.14^{+0.53}_{-0.58}$&  0.45$^{+0.06}_{-0.06}$ & 0.06 & 0.03 \\
2173 & LMC &  Clay(2/11) & 10800 &  9.33 & 4.87 &$-$0.24 &  9.6 & 2.11 & $3.39^{+0.21}_{-0.22}$&  1.66$^{+0.21}_{-0.21}$ & 1.36 & 0.76 \\
2213 & LMC &  Clay(2/11) &  9000 &  9.20 & 4.53 &$-$0.01 &  4.3 & 2.46 & $3.54^{+0.33}_{-0.36}$&  1.79$^{+0.35}_{-0.34}$ & 1.14 & 0.64 \\
2249 & LMC &  Clay(2/11) & 14400 &  8.82 & 4.63 &$-$0.47 &  3.7 & 2.69 & $3.01^{+0.36}_{-0.40}$&  0.88$^{+0.22}_{-0.22}$ & 0.46 & 0.26 \\
4590 &  MW &  Clay(2/11) &  3600   & 10.11 & 4.69 &$-$2.06 &  4.3 & 2.63 & $3.18^{+0.26}_{-0.27}$&  0.98$^{+0.17}_{-0.16}$ & 3.13 & 1.89 \\
\enddata
\label{tab:clusters}
\end{deluxetable*}

\subsection{Integrated Spectra}

We obtained our spectra using the MIKE spectrograph \citep{mike} at the Magellan II (Clay) telescope during Nov. 2006, Aug. 2009, Feb. 2011, and Oct. 2011, and with the echelle spectrograph at the 100-inch at Las Campanas (du Pont) during Dec. 2000 and Jan. 2001. 

The MIKE spectrograph provides blue and red channel spectra over the entire optical window. We use the data from the red side, in which the most prominent absorption lines with high S/N lie, and confine ourselves to shortward of 7000 \AA\ to avoid strong atmospheric features. We used the 0.75 arcsec slit to avoid degrading resolution further, which is already near the limit for clusters that can possibly have $\sigma \sim 2$ km s$^{-1}$. The exposure times vary depending on the surface brightness and are given in Table \ref{tab:clusters}. 

To obtain integrated spectra of the clusters, we utilize the same spectroscopic drift technique described by \cite{colucci1}. To summarize, we set the telescope in motion to raster scan the slit across the cluster during the exposure, defining both the angular length and height of the raster (both set to the same number). The exposure time then sets the rate of the scan, such that the full scan is completed within the allotted exposure time. We chose from only two different values of the scan sizes (10\arcsec $\times$ 10\arcsec or 30\arcsec $\times$ 30\arcsec), aiming for the closest match to the half light radius of the cluster. Because MIKE does not have an instrument rotator, these scans are then further complicated by field rotation, which is unaccounted for. Multiple exposures further help homogenize the sampling of the central region. 
Our goal is to obtain a spectrum that is representative of the central region of each cluster rather than one of a tightly specified region.
The slit is only 5 arcsec long, so these raster scans have no clear sky. We do not attempt sky subtraction. Solar spectral features, from moonlight, are at significantly different velocities and are also quite weak as the two key runs (the last two) were both in dark time (moon illumination $<$ 30\%). 

We reduced the MIKE data with the MIKE DR pipeline (R. A. Bernstein, in preparation, available at http://www.ucolick.org/$\sim$xavier/Profession.html).
Because of the scanning technique, the cluster light fills the slit, so that a modification to the pipeline was developed that allows boxcar rather than optimal source extraction.  The data obtained in Aug., Feb. and Oct. 2011 generally have shorter exposure times, fewer frames and more intrinsic emission lines from the youngest clusters, which render the standard cosmic ray rejections in the MIKE DR and IRAF\footnote{IRAF is distributed by the National
  Optical Astronomy Observatories, which are operated by the
  Association of Universities for Research in Astronomy, Inc., under
  cooperative agreement with the National Science Foundation.}  packages ineffective for producing the combined spectra.  For these data, we remove the cosmic ray and emission spikes in the individual frames using a 10-sigma threshold above a continuum fit and median combine the cleaned frames using the IRAF $scombine$ routine.  Again, we remove the blaze function using a trace of the continuum flux of a G-type star.

A second set of data come from observations with the du Pont telescope and its echelle spectrograph. Those spectra have a wavelength coverage of approximately 3700 to 7800 \AA, with declining sensitivity and spectral resolution toward the blue end.  These spectra were also obtained using the scanning technique.  At the du Pont telescope scanning was implemented using a modification of the telescope guider program provided by S. Shectman (see McWilliam \& Bernstein, 2008). The echelle slit is 1\arcsec x 4\arcsec,  allowing uniform coverage of a 12 $\times$ 12 arcsec$^2$ or 8 $\times$ 8 arcsec$^2$ high-surface brightness region of the cluster. We took multiple exposures to homogenize the scanned cluster region and for cosmic ray removal. We reduced the spectra using standard IRAF routines (see Colucci et al. 2011), including the scattered light subtraction described in detail in McWilliam \& Bernstein (2008). We combined the extracted spectra using the IRAF $scombine$ routine with the $crreject$ algorithm to eliminate cosmic ray events. Finally, we remove the blaze function using a trace of the continuum flux of a G-type star.

\subsection{Measuring Velocity Dispersions}

Various techniques have been developed and applied to the problem of measuring velocity dispersions from integrated spectra \citep[for example,][]{sargent,bender,rix,kuijken,winsall}. The appropriate approach depends on the particulars of the scientific situation and data quality. In our case, we have good S/N for most of our clusters, and therefore the luxury of not having to rely on a method that utilizes the entire spectrum at one time so as to increase the S/N, but which incurs the cost of obscuring potential systematic errors. Specifically, we have enough S/N that one trustworthy line would be sufficient to provide a results with the desired precision. As such, our primary concern is systematic errors.

Systematic errors in this type of measurement can arise from a variety of sources including differential instrumental broadening across the spectra, template mismatch, errors in wavelength calibration, and errors in the definition of the continuum level. These are all much easier to identify if one can work with individual absorption lines rather than with a vaguely weighted mean from the combination of all lines. Some of these errors can be ameliorated by fitting on a case-by-case basis, for example one can let the radial velocity float in the fits of individual lines to mitigate against wavelength calibration errors, minimizing the risk that an inflated $\sigma$ will be inferred as the algorithm attempts to better fit a number of lines with a single radial velocity. Other errors, such as those that arise for lines near the ends of orders,  will result in ``problem lines" being identified as outliers in the set of measurements. For these reasons, we choose to fit as many of the absorption lines as we can individually and then use the entire set of measurements to identify the most likely value of the velocity dispersion. 

Our selection of lines to be fit is based on a visual inspection of a long list of lines, originally identified in our template star spectra. We reject any lines in the object spectra that are either clearly blends or suffer some other complication and those that do not dip below 0.85 of the local continuum level.  After fitting, we reject lines that do not produce an acceptable fit, where acceptable is defined by $\chi^2 < 2.3$. To calculate $\chi^2$ we adopt a per pixel uncertainty determined from the fluctuations about a flat continuum in line-free areas of the spectrum. However, we adopt the same uncertainty value for the full spectral range for any given cluster. Our results are not highly sensitive to this value because we only use these $\chi^2$ values to remove questionable fits from further consideration, and visual inspection confirms that those lines that have been rejected by this criteria are poorly fit. 

\begin{figure}[htbp]
\plotone{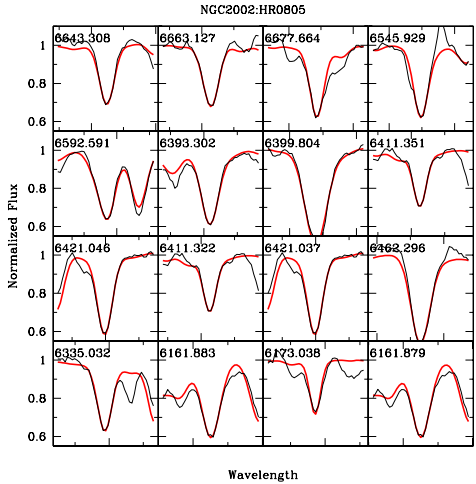}
\caption{Example absorption line fits drawn from data for NGC 2002, using stellar template HR 805. Each line is fit independently using a Gaussian convolution kernel and scaling for both slightly different radial velocities and equivalent width. Best fits are shown with the smooth line (red). These are a subset of the absorption lines available for NGC 2002. }
\label{fig:examples}
\end{figure} 

\begin{figure*}[htbp]
\plotone{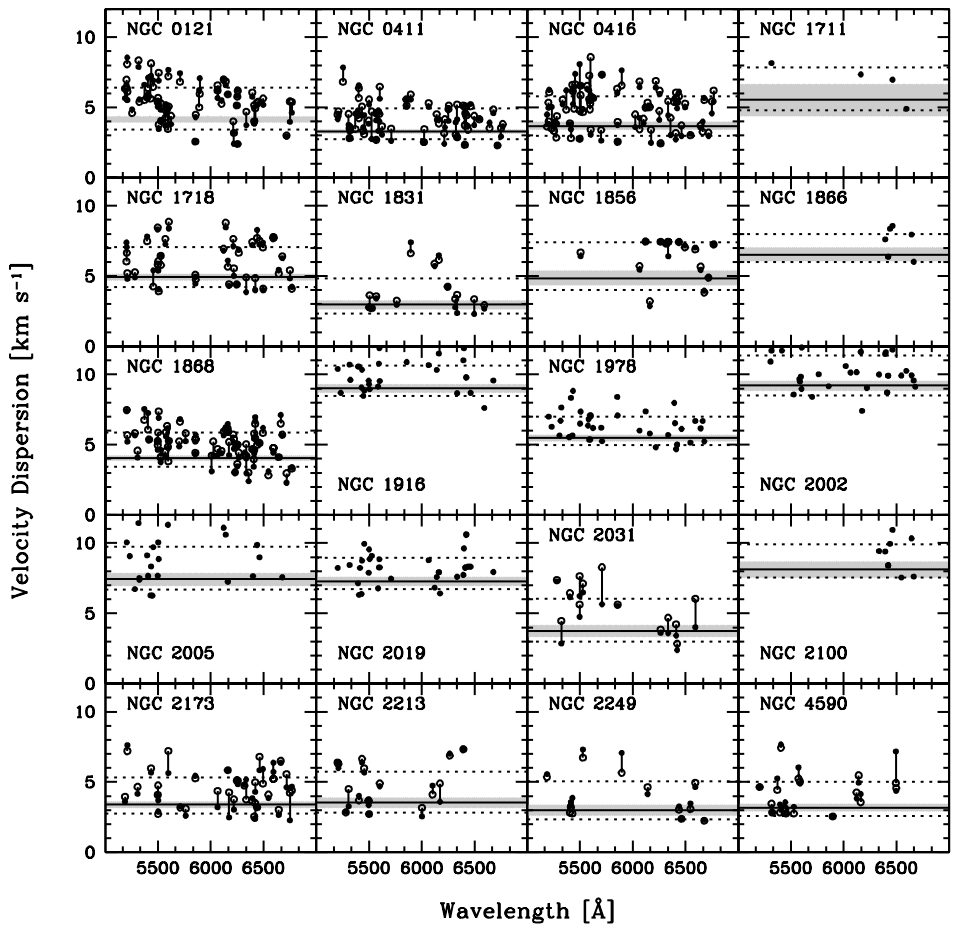}
\caption{Results from the velocity dispersion determination. In each panel we present the results from our measurement of velocity dispersions using individual absorption lines that passed our selection criteria for a different cluster. In cases where we also compare results using different template stars, we use filled and open circles and connect the results with vertical lines. We find no significant systematic difference in $\sigma$ arising from our choice of template. The best fit dispersion value for each cluster is shown as the solid line, with the grey area surrounding it demarcating our 90\% confidence interval on this mean value. The dashed lines show the uncertainty, derived as explained in the text, for an individual measurement. }
\label{fig:figall}
\end{figure*}

Some of the subtleties involved in selecting among the various methods in the literature to measure $\sigma$ lie in the possible nature of the underlying line-of-sight velocity distribution (LOSVD), which can be quite complex in galaxies. 
It is our expectation that in clusters, with relatively short dynamical times, the LOSVD is nearly Gaussian. As such, our fitting involves the convolution of a Gaussian, which we adopt as the description of the LOSVD, of specified width with the rest frame template spectrum. For our MIKE cluster spectra, we use two template stars and later compare results (we find no significant differences in the inferred $\sigma$). The convolved spectrum is then redshifted by a specified value corresponding to a selected radial velocity, $v_r$, and binned into the same wavelength bins as the object spectrum using cubic splines. Finally, we also explore a range of equivalent width renormalizations so that we can match the strength of the line in the template to that in the object spectrum. The renormalization is done by subtracting 1, the value of the normalized continuum, from the spectrum, multiplying the spectrum by a specified factor, and adding the 1 back to the spectrum. We sample the multidimensional parameter space ($\sigma$, $v_r$, normalization) uniformly and in an unbiased way,  calculate $\chi^2$ for $\pm 3$ pixels about the line center  (in a few cases where the spectral lines are narrow we set this to $\pm 2$ pixels) because we do not want the selection of the best fit parameters influenced by nearby spectral lines or slight continuum mismatches, and identify the best fit parameters. The focus on the core of the line minimized various systematic technical problems, such as fluctuations in the continuum, but also addresses some external sources of uncertainty, such as the role of binaries (\S 4) and interlopers, by placing less weight on the wings of the line-of-sight velocity distribution. Examples of fits for absorption lines in the spectra of NGC 2002 are provided in Figure \ref{fig:examples}.

We correct for instrumental broadening using the template star spectra. Each of the absorption lines we identify in the template spectrum is fitted with a Gaussian. Assuming that the intrinsic line widths are well below the instrumental broadening, which is in the range of 2 km s$^{-1}$, then the fitted Gaussian widths as a function of wavelength describe the instrumental broadening. We do indeed find a strong correlation between the Gaussian width in km s$^{-1}$ and wavelength, as expected from instrumental broadening.  We fit a low order polynomial to that relation and use it to estimate the instrumental broadening at any wavelength. This procedure is done independently for each run and, of course, each of the two instruments used. The inferred instrumental broadening is added in quadrature to the best fit value of $\sigma$ derived above because we have fit a template line that has already been broadened.

Once the line fitting is complete, the question becomes how to use the values derived for the set of individual absorption lines to best derive the value of $\sigma$. The use of individual lines, even when visually vetted, does not eliminate the potential for systematic effects that inflate the widths (such as blends and focus errors). 
The internal errors provided by the fits themselves are usually optimistic given the high S/N of these data. 
We evaluate the final value of $\sigma$ for each cluster by calculating the average of all the measurements 
and $\chi^2/N$, where $N$ is the number of data points.  We set the uncertainty in any individual measurement by requiring $\chi^2/N = 1$. To downweight measurements that are inflated by blends or focus errors, we set upward uncertainties to be $X$ times larger than downward ones. 
We will discuss the selection of $X$ shortly. The uncertainty on our final ``mean" $\sigma$ is  derived by identifying the range of $\sigma$'s that generate $\chi^2 - \chi^2_{min} < 2.71$, corresponding to the 90\% confidence interval (see Table \ref{tab:clusters}).

Different weighting factors $X$ will result in different values of $\sigma$. As such, without further constraints on $X$, this parameter is itself a source of systematic uncertainty, and simply reflects the uncertainty associated with unknown blends, focus errors, etc. 
To the degree that changing $X$ results in similar proportional shifts to all the $\sigma$'s, this effect only results in global shifts in the values of $\Upsilon_*$, whose normalization is uncertain for other reasons as well. Nevertheless, for guidance we examine the internal uncertainty on the individual fitted lines produced by our fitting algorithm, which have upward uncertainties that are between 1.5 and 3.5 times larger than downward uncertainties. For our calculations, we therefore select a value within that range, $X = 3$, and note that choosing $X$ anywhere in this range
produces resulting $\sigma$'s that lie within the quoted uncertainties in Table \ref{tab:clusters}.

In certain panels in Figure \ref{fig:figall}, pairs of points are connected with lines. These points and lines represent two measurements of the same absorption line using different template stars. 
We find that on average, over the twelve clusters for which we used two templates, the velocities of individual lines are biased by $0.12$ km s$^{-1}$ from one template star to another --- a value smaller than any of  the internal uncertainties we quote. Therefore, we conclude that template selection does not have a dominant effect on our uncertainties. 

\begin{figure}[htbp]
\plotone{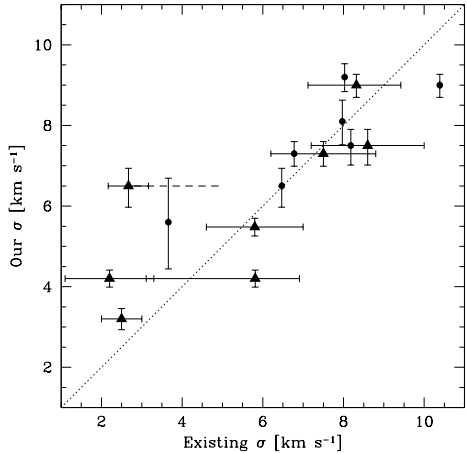}
\caption{A comparison of our measurement of $\sigma$ to two other measurements. First, we compare to determinations using our data but a different technique (cross-correlation) in the filled circles. Second, we compare to existing measurements in the literature \citep{mclaughlin} in the filled triangles. The line represents the 1:1 relation.}
\label{fig:sigcomp}
\end{figure}

For the subset of our clusters that have been observed previously, we compare our $\sigma$ measurements to pre-existing values. First, one of us (JC) has independently (although using the same spectra) obtained estimates of $\sigma$ using a cross-correlation analysis. This provides an internal test of our methodology. Second, for an external test, we turn to the compilation of \cite{mclaughlin}, which includes literature values of $\sigma$ for six of our clusters, and other sources
that include two more \citep{meylan,lane}. Both internal and external tests
are
presented in Figure \ref{fig:sigcomp}. 
The external comparison is positive in that 5 of 8 (63\%) of the measurements agree to 1$\sigma$ and 
7 of 8 (88\%) agree to 2$\sigma$. For the one 
significant outlier
NGC 1866, 
the \cite{lane} study can be used to demonstrate the sensitivity of that particular result, obtained using individually measured stars, to the inclusion or exclusion of single star (dashed line in Figure \ref{fig:sigcomp}).  

There are two clusters, NGC 458 and 1860, for which we were unable to identify any suitable lines for measurements. This is somewhat surprising given exposure times that are in the range of those obtained for other clusters. NGC 1860 is particularly large, nearly twice as large as the next nearest cluster in size, and has a correspondingly low mean surface brightness, which may have just been too low for these observations. The situation for NGC 458 is a bit more puzzling, but we may have simply been unfortunate in the region we scanned. 

\section{Determining Masses}

The single-epoch dynamical determination of the mass of an astronomical system always references the Virial Theorem. In practice, the different methods that have been developed are all attempts to deal with the unknown numerical factors arising from the integration of the mass distribution and kinematics required when calculating the terms in the Virial Theorem. In certain cases, where the number of independent tracer particles with measured velocities is small, there is a concern for numerical stability of the mass estimator \citep[cf.][]{bahcall}, but usually the concern is how to properly weight the kinematic term, related to $\sigma$, given the unknown distribution of tracer particle orbits. Theoretically, the difference between the mass inferred for a system that has a tangential distribution of orbits rather than a radial one can be as large as a factor of three if the incorrect orbital anisotropy is assumed. There are observational ways to constrain the orbital anisotropy using higher order measures of the line of sight velocity distribution \citep[see][]{winsall,kuijken}, but such analyses are beyond
the scope of the current work.

A surprising work-around to this problem has been identified recently by \cite{walker} and \cite{wolf}. They find that the mass enclosed at the half light radius of stellar systems, and only the half light radius, is nearly insensitive to such details as the orbital anisotropy or the radial profile of the tracer population. They have gone on to use this finding to estimate the masses of dwarf spheroidal galaxies in particular \citep{walker}, but also of other spheroidal stellar systems \citep{wolf}.  By examining a range of dynamical models \cite{walker} found  a robust estimator for $M_h$ to be $M_h = 580r_h\sigma^2$, where $M$ is in solar masses, $r_h$ is in parsecs, and $\sigma$ is in km s$^{-1}$. 

Such an estimator is even more trustworthy if verified empirically. Of particular importance is whether the estimator is valid over the full range of system masses and the degree to which mass estimates for individual systems fluctuate about the mean value described by the estimator.
Interestingly,  the \cite{walker} estimator, when rewritten in the style of the Fundamental Manifold \citep{z08}, is equivalent to 
\begin{equation}
\log r_h = 2\log \sigma - \log I_h - \log \Upsilon_h - 0.73,
\label{eq:walker}
\end{equation}
where now $r_h$ is given in kpc, $\sigma$ is still in km s$^{-1}$, $I_h$, which is the mean surface brightness within $r_h$ is in solar luminosities per sq. parsec, and $\Upsilon_h$ is the mass-to-light ratio within $r_h$. 
By measuring $r_h$, $\sigma$, and, $I_h$, one can solve for $\Upsilon_h$. 
\cite{z08,z11} identified an almost identical empirical scaling relationship, independent of the \cite{walker} study, that works for all stellar systems, ranging from the most massive galaxies to globular clusters. They expressed their relationship as
\begin{equation}
\log r_h = 2\log \sigma - \log I_h - \log \Upsilon_h - 0.75.
\label{eq:fm}
\end{equation}
where the zero point, the 0.75 term, comes from placing galaxies with independent mass estimates  from detailed dynamical modeling \citep{cappellari} on the relationship. As is already evident, this empirical calibration produces an almost exact match in the normalization (0.73 vs 0.75) to that provided by the theoretical modeling of \cite{walker}. 
The empirical results verify that there is little scatter ($\sim 0.1$ dex) about this relationship for objects ranging from globular clusters to massive elliptical galaxies.
Using Equation \ref{eq:fm}, we evaluate $\Upsilon_h$ for our set of 20 clusters and present the results in Table \ref{tab:clusters}. 
Masses can be calculated using $\Upsilon_h$ and the total luminosities, which we also provide in the Table. For systems without dark matter, which we presume includes these clusters, $\Upsilon_* \equiv \Upsilon_h$. All photometric quantities are presented for the $V$ band.

The uncertainties in $\Upsilon_*$ are calculated using only the uncertainty in $\sigma$, as the internal uncertainties on the other parameters are proportionally much smaller. However, we test for the possible systematic effect in $r_h$ and $I_h$ due to our choice of the Wilson model fits by comparing estimates of $\Upsilon_*$ using the values given by \cite{mclaughlin} for their alternate King and power-law fits. We find that with the exception of two clusters (NGC1856 and NGC2100), where the power-law fit gives significantly different estimates of $r_h$ and $I_h$, all estimates of $\Upsilon_h$ are within $2\sigma$ and 80\% are within 1$\sigma$. For the two discrepant clusters, the power-law fit to NGC1856 results in an unphysically large value of $r_h$, and hence $\Upsilon_*$, and in both cases the results of the King and Wilson models agree within the uncertainties.

\subsection{Comparison to Other Estimates of $\Upsilon_h$}
\label{sec:comp}

\begin{figure}[htbp]
\plotone{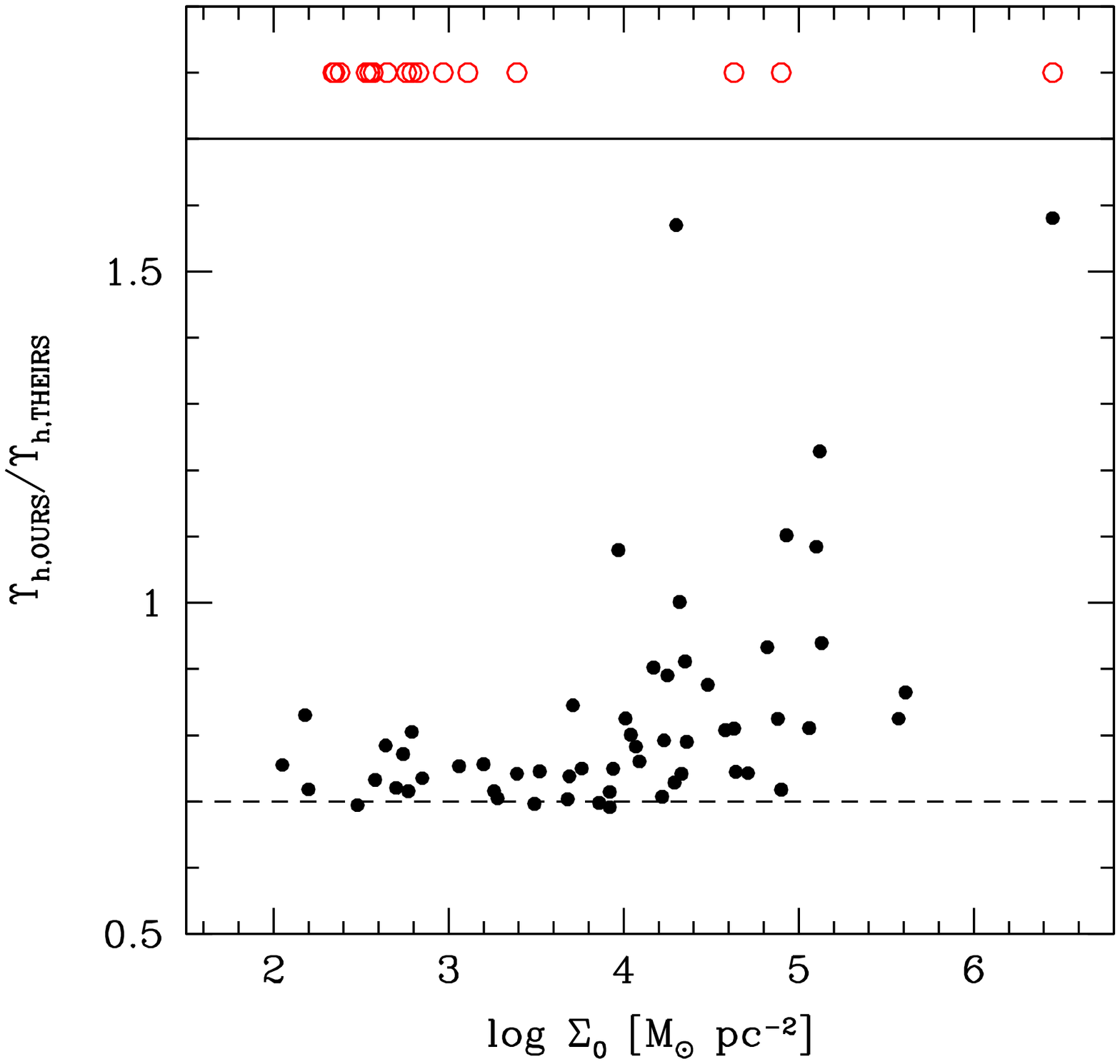}
\caption{The dependency of the ratio of $\Upsilon_h$ estimates from \cite{mclaughlin} (THEIRS) to our own (OURS) on central projected mass density. The bottom portion of the Figure illustrates how the our estimate of $\Upsilon_h$ is well behaved relative to that provided by \cite{mclaughlin} for central projected mass densities that are $< 10^4 M_\odot$. There is a systematic offset where our values tend to be lower by $\sim$ 30\% (the dashed lines shows a ratio of 0.7 for reference). Above that value there is large scatter, with an implication that our method could produce significant overestimates of $\Upsilon_h$ for some large values of the central mass surface density. The top portion of the Figure illustrates the distribution of our sample with respect to central projected mass density. Only three of our clusters lie in the suspect region, although those three are all old (log(age) $>$ 10) clusters and our estimates of $\Upsilon_*$ for these are low.}
\label{fig:sb}
\end{figure}

We compare our estimates of $\Upsilon_h$ using Equation \ref{eq:fm} with those obtained by \cite{mclaughlin}. Using the published values of $\sigma$, but our mass estimator, we obtain values of $\Upsilon_*$ that tend to be lower than those obtained by \cite{mclaughlin} by $\le$ 30\% (Figure \ref{fig:sb}). However, global shifts of this magnitude are expected when using different estimators due to the difficulty in determining the zero point calibration of such estimators. More importantly, the clustering of points about a fixed ratio of the two estimators shows that the correlation between the two is excellent. If we were to recalibrate Equation \ref{eq:fm} to give consistent answers, we would change the calibration constant in the Equation from 0.75 to 0.60. We will leave the constant as is, partly due to the agreement with the results from \cite{walker}, but appreciate that there is a degree of freedom available in the overall normalization of the values of $\Upsilon_h$. That freedom is a global shift in $\Upsilon_h$ rather than one that depends on $\Upsilon_h$ itself. 

Examining the differences in detail (Figure \ref{fig:sb}),  we find that there is a marked connection between the difference in the values of $\Upsilon_h$ for individual clusters and their central projected mass density (and related quantities like volume density and phase space density). Below central mass surface densities of $10^4$ M$_\odot$ pc$^{-2}$, the two mass estimators track each other well, modulo this $\le$ 30\% offset. For central mass densities greater than this, the scatter increases significantly. This does not, in itself, suggest which of the two approaches should be preferred, but suggests some caution in treating estimates of $\Upsilon_h$ for high density clusters. In our sample, only three clusters (NGC 1916, 2005, and 2019; see red, open symbols in the plot) lie above this threshold. We find no connection between the differences in mass estimates and other quantities provided by \cite{mclaughlin}, such as relaxation time at the half mass radius or total mass.
The discrepancy, if the \cite{mclaughlin} estimates are the ones to be trusted, is in the sense that we would be overestimating $\Upsilon_h$ for some fraction of high central density clusters. The three clusters in our sample that are potentially at risk are all old clusters (log(age [yrs])$>$10), but our estimates of $\Upsilon_h$ and those of \cite{mclaughlin} agree to within 5\% (after adjusting for the 30\% mean offset). 

\section{Stellar Mass-to-Light Ratios}

\begin{figure}[htbp]
\begin{center}
\plotone{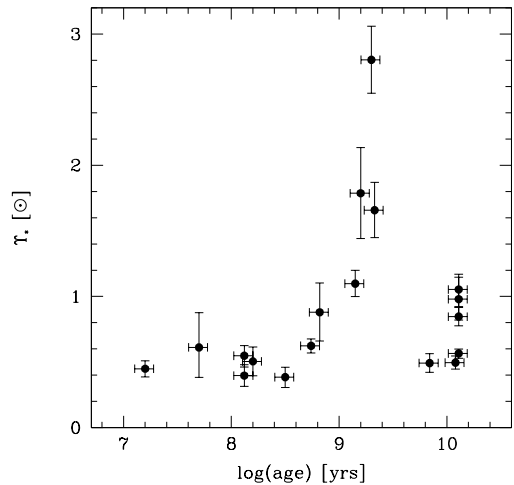}
\caption{Stellar mass-to-light ratio, $\Upsilon_*$ versus age. We plot the dynamical estimates of $\Upsilon_*$ for our homogeneous sample of 18 stellar clusters. }
\label{fig:m2l}
\end{center}
\end{figure}

Using the derived values of $\Upsilon_h$ and assuming that the clusters are devoid of exotic dark matter, we show our principal empirical result, the relationship between $\Upsilon_*$ and age, in Figure \ref{fig:m2l}. Here we see a definite trend, where $\Upsilon_*$ rises as we consider older and older clusters up to ages of a few Gyr, and then drops significantly for the oldest clusters. 
While this drop runs counter to the naive expectation that clusters should continue to fade with age and therefore that $\Upsilon_*$ should continue to rise, the evaporation of low mass stars due to two-body relaxation will alter this expectation, particularly at the oldest ages, as discussed in \S4.1.

Before continuing to explore all of this further, we note that
the youngest two clusters in our sample are close to or below the 50 Myr age for which \cite{goodwin} argue that dynamical mass estimates are unreliable because clusters are still undergoing violent relaxation. They show that dynamical masses can be inflated by factors of several, although these effects go away for log(age [yrs])$ > 8$. We therefore place no weight on disagreements found between observations and models at the youngest ages (log(age [yrs]) $< 8$.

\begin{figure}[htbp]
\begin{center}
\plotone{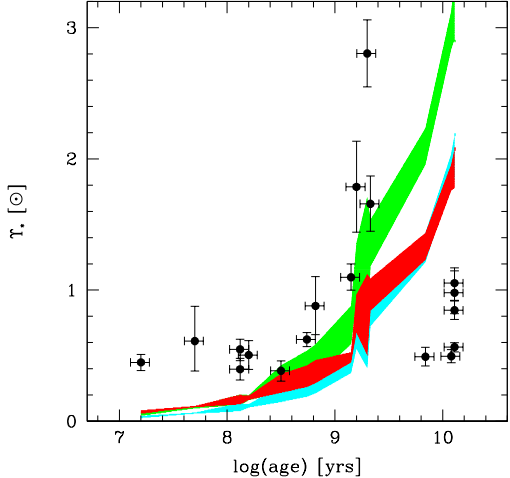}
\caption{Stellar mass-to-light ratio, $\Upsilon_*$, versus age compared to model prediction. To Figure \ref{fig:m2l} we have now added the calculated values of $\Upsilon_*$ for different stellar population models as evaluated by \cite{mclaughlin}. The upper (green) curve represents the results obtained using the \cite{bc} algorithm and a \cite{salpeter} IMF. The intermediate (red) represents results using the same algorithm by a \cite{chabrier} disk IMF. Finally, the lowest curve represents results obtained using the \cite{pegase} algorithm and the \cite{kroupa} IMF. All three fail in quantitative detail to fit the data. The clusters with log(age) $\le$ 8 may not be relaxed and are not given any weight in this comparison. Dynamical effects that may affect the older clusters are discussed in the text, but found to be insufficient to reconcile the data and models.}
\label{fig:m2lmodel}
\end{center}
\end{figure} 

We compare our results to models of simple stellar populations (Figure \ref{fig:m2lmodel}). Specifically, we begin with the results presented by \cite{mclaughlin} using either the \cite{bc} or PEGASE \citep{pegase} models and either \cite{salpeter}, \cite{chabrier}, or \cite{kroupa} IMFs. Estimates of $\Upsilon_*$ were calculated using the metallicities listed in Table \ref{tab:clusters}, and so in theory account for the different metallicities of the old and young clusters. The models clearly straddle the data (Figure \ref{fig:m2lmodel}) but fail both by underestimating $\Upsilon_*$ for the younger clusters and overestimating it for the older clusters. Simple multiplicative rescaling of $\Upsilon_*$ is allowed because our normalization (Equation \ref{eq:fm}) could be questioned --- for example, our calibration based on elliptical galaxies might not be an exact match for stellar clusters --- but a simple multiplicative shift  of the $\Upsilon_*$ measurements cannot address an underprediction at young ages and an overprediction at older ages. 
The only way to reconcile the data and models is to posit that either the observations or the stellar population models have a systematic error that depends on age (or a related parameters such as metallicity). 

One might suspect the results because there are only a handful of points at the oldest ages. Perhaps these are anomalous among older clusters. The sample provided by \cite{mclaughlin} contains 51 old clusters (log(age [yrs]) $>$ 10) and the median of the $\Upsilon_*$ distribution for those clusters, when estimating $\Upsilon_*$ with Equation \ref{eq:fm}, is 1.2, or about 40\% larger than the average $\Upsilon_*$ for our five old systems (0.85). This difference includes the 30\% offset discussed in the previous section that is the result of a zero point mass offset, so the additional factor is small.
We conclude that while our clusters may have somewhat unusually low values of $\Upsilon_*$, the likely mean value still lies well below the peak $\Upsilon_*$ seen at intermediate ages, and certainly does not lie on the rising extrapolation of the relation seen for the younger clusters.  

There are, of course, physical reasons why the results may be distorted.
First, the older clusters are all of significant lower chemical abundance than the younger clusters (see Table \ref{tab:clusters}). This is certainly a source of concern, although we are comparing the values of $\Upsilon_*$ to those predicted by models that account for metallicity differences.  Second, these clusters could, for some unknown reason, not satisfy the mass estimator of Equation \ref{eq:fm}. Aside from the high central projected surface brightnesses of three of these, there is no striking difference in structural parameters. They tend to be among the physically most compact, most concentrated ones, but not exclusively so (NGC 121 and NGC 4590, fall within the range of the majority of the clusters). Furthermore, several of these (NGC 1916, 2005, 2019, and 4590) are also among the clusters studied by \cite{mclaughlin} and the mean ratio of our $\Upsilon_*$ estimates to theirs is 0.6. Recall that for the entire sample of old clusters this ratio is $\sim$0.7, so we see no unusual behavior in the estimated $\Upsilon_*$ values for our set of old clusters beyond the increased scatter at high surface brightnesses that was discussed previously. If anything, the comparison to the \cite{mclaughlin} results, which highlighted the issue with high surface brightness clusters, suggests that we are likely to be overestimating $\Upsilon_*$ for this set of clusters. Disregarding this comparison and hypothesizing 
a factor of two upward scaling error in our $\Upsilon_*$ values would allow our cluster $\Upsilon_*$ estimates to come to better agreement with the evolutionary models that use Chabrier or Kroupa IMFs, but would exacerbate the underprediction of $\Upsilon_*$ for the intermediate age clusters. We find
no evident reason why the data for the older clusters in our sample should be singled out as susceptible to calibration  biases in $\Upsilon_*$ by factors $>$ 2.

\begin{figure}
\begin{center}
\plotone{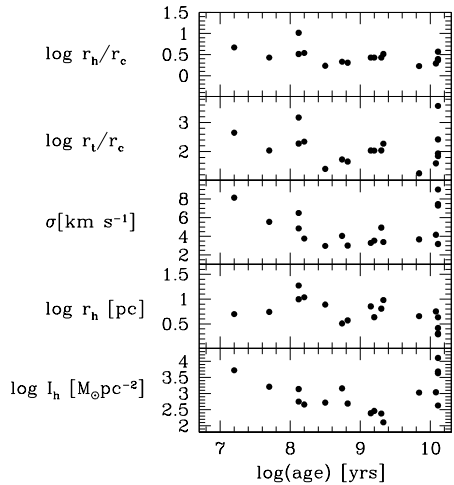}
\caption{Cluster structural properties as a function of age. The old and intermediate age clusters do not have strikingly different structural properties.}
\label{fig:comp}
\end{center}
\end{figure}

Alternatively, one could be suspicious of the intermediate age clusters. However, the values of $\sigma$, $r_h$, and $I_h$ for these clusters are roughly within the value ranges for the other clusters (Figure \ref{fig:comp}), with the possible exception that their surface brightnesses are somewhat lower. The latter is due to the lack of intermediate age clusters available for study and our selection for the highest surface brightness clusters available.  Even so, this range of surface brightnesses is not outside of the range of our comparison to the modeling results of \cite{mclaughlin} (Figure \ref{fig:sb}) and so we do not see this as a source of concern.

The only way to reconcile the data and models is to posit that either the observations or the stellar population models have a systematic error that depends on age (or a related parameters such as metallicity). We have just argued against systematic errors in our sample selection or measurements, and so we are left with the possibility of age-dependent problems in the application of simple stellar population or dynamical models to predict $\Upsilon_*$. We address two potential reasons for such failures below.

\subsection{Dynamical Evolution of Clusters and Its Impact on $\Upsilon_*$}
\label{sec:dynamical}

When using stellar clusters, particularly as a test of stellar population models, one needs to account for dynamical effects that lead to the preferential loss of low-mass stars and alter the simple predictions for $\Upsilon_*$ shown in Figure \ref{fig:m2lmodel}. Clusters can be significantly affected by mass loss driven by gas removal at early times \citep[for examples see][]{hills,goodwin}, and by stellar evolution in a tidal field, gravitational perturbations due to passing molecular clouds, ejections during binary interactions, and evaporation of low mass stars via two-body interactions 
\citep{spitzer, kruijssen08} on longer timescales. 
Modern treatments of the evolution of an isolated cluster or one in a tidal field are presented by \cite{gieles10,gieles11}.  
We concentrate on the evaporation of low-mass stars via two-body interactions, which has the strongest impact on the oldest clusters, to determine the degree to which the measured values of $\Upsilon_*$ have been lowered relative to the simple stellar populations expectations.

\begin{figure}[htbp]
\begin{center}
\epsscale{0.9}
\plotone{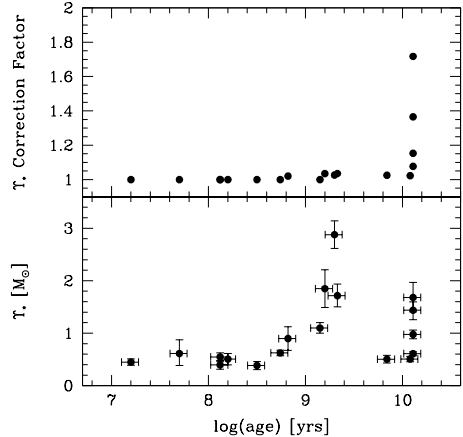}
\caption{Correcting for the effects of dynamical evolution on $\Upsilon_*$. In the upper panel we show the results from applying the models of \cite{anders}, which address the issue of dynamical evolution of the clusters. The correction is in the sense of accounting for lost low mass stars from clusters so as to provide a fair comparison to simple stellar population models.
 We plot the ratio of $\Upsilon_*$ obtained from a model with effectively no evolution, setting $t_4 = 100$ Gyr, to ones where we have taken the lowest, self-consistent values of $t_4$ (see text for details). The dynamically derived values of $\Upsilon_*$ would need to be multiplied by the plotted value to obtain an estimate of $\Upsilon_*$ for a population with no lost stars. In the lower panel, we apply that correction and find that although there is an increase in the values of $\Upsilon_*$ for the older clusters, the qualitative nature of the behavior of $\Upsilon_*$ with age remains unchanged and will still not match the expectations drawn from simple stellar population models.}
\label{fig:anders}
\end{center}
\end{figure} 

Low-mass clusters are disrupted earlier than high-mass clusters of similar
density due to internal two-body relaxation.
The evaporation rates depend on a number of poorly constrained parameters, such as the internal density profile and tidal field strength. It is therefore difficult
to predict the evaporation rate
for any particular cluster.  Instead, the luminosity and 
mass function of globular clusters provides
constraints on the evaporation rate for populations of clusters.
The mass function of globular clusters in the Milky Way has a peak
near $\approx 1-2 \times 10^{5}~M_{\odot}$.
Fall \& Zhang (2001) dynamically evolved simulated globular
cluster systems and showed that this peak results from the earlier
disruption of low-mass clusters relative to high-mass clusters of
similar density, where the mass of each cluster is depleted
approximately linearly with time, with an evaporation rate
$\mu_{ev} \approx 1-2 \times 10^{-5}~M_{\odot}~\mbox{yr}^{-1}$.
The mass function of globular clusters in the LMC has a similar shape
to those in the Milky Way, although the uncertainties on this shape
are larger because of lower numbers (Chandar et al. 2010). Nevertheless, 
this result suggests that the ancient globulars in our LMC sample have experienced
similar evaporation rates, on average, as those in the Milky Way.

Various authors have studied the impact of relaxation driven evaporation on $\Upsilon_*$ in star clusters.
For example, Lamers et al. (2005a) and Anders et al. (2009)
used N-body simulations of star clusters to predict total
cluster lifetimes and the evolution of 
$\Upsilon_*$ for clusters with different stellar mass-loss rates.
Relative to standard models, low-mass stars, which have a higher $\Upsilon$ than 
that of the typical cluster star, are lost, but the fraction of non-luminous 
stellar remnants increases and these two effects partially cancel.
Anders et al. (2009) present calculations for the disruption time for clusters, $t_d$, which 
depends on the cluster
mass and local (external) density, with a mass dependence of
$t_d \propto M^{0.62}$. This relationship is typically normalized by specifying the time
at which a 10$^4$ M$_\odot$ cluster loses 95\% of its mass, $t_4$.
The mass dependence can be converted into a mass loss rate $dM/dt \propto M^{0.38}$,
which is shallower than the relation given by classical evaporation and
gives a poor fit to the overall 
shape of the
mass function of globular clusters in M87 and in the Milky Way \citep{waters}, particularly at the low mass end.
Nevertheless, these models can reproduce the peak of the mass function. In principle, 
each cluster has its own corresponding value of
$t_4$ depending on its internal density, orbit, and tidal field
strength, although we have insufficient information to evaluate
$t_4$ on an cluster-by-cluster basis, and a single value of $t_4$ is usually assumed for a set of clusters.

To evaluate the magnitude of the effect on $\Upsilon_*$, we choose from among the \cite{anders} models for a selected value of  $t_4$ (see below) and the nearest values of  $\langle$Fe/H$ \rangle$ and age for the particular cluster. In Figure \ref{fig:anders} we show the ratio between the predicted values of $\Upsilon_*$ from these models to those obtained from models with effectively no evolution ($t_{4} \sim$ 100 Gyr). 
To select values of $t_4$ we adopted values from the literature. In the SMC, $t_{4}$ is estimated to be roughly 10 Gyr \citep{lamers05b}, while in the local neighborhood it is estimated to be between 1.3 Gyr \citep{lamers05a} and $\sim$5 to 6 Gyr \citep{baumgardt,lamers05b}. We adopt $t_{4} = 10$ Gyr for clusters in the SMC and then scale by mass for our SMC clusters. We adopt $t_{4}$ = 5 Gyr for the remaining clusters, which is the lowest possible value of $t_{4}$ that produces self-consistent results in the sense that none of our clusters are older than the inferred dissolution time. The results shown in Figure \ref{fig:anders} demonstrate that the effect is at most a factor of two increase in $\Upsilon_*$ although usually significantly less. Hereafter, our values of $\Upsilon_*$ are the corrected values.
A comparison between the present-day stellar mass functions measured
for 27 old globular clusters and simulations of cluster evolution suggest
that the present-day mass functions are consistent with expectations
from two-body relaxation for a given universal IMF \citep{leigh}.

There is, of course, uncertainty in the selection of $t_4$. However, there are two reasons why we conclude that dynamical evolution cannot reconcile the models to the data even if one is allowed to change $t_4$ on a cluster-by-cluster basis. 
First, the effects of dynamical evolution are at best only modest ($\sim$ 2) and reconciling the data to the better fitting 
Salpeter model (Figure \ref{fig:m2lmodel}) requires significantly larger corrections for the entire sample of old clusters. Second, despite the choice of $t_4$ within reason, some of these old clusters are sufficiently massive that they are impervious to the effects of dynamical evolution.

\section{Binaries}

If the old clusters are behaving as expected, then the problem might lie with the intermediate age clusters. If so, one possibility is that their velocity dispersions are inflated by binaries. 
Contamination by binaries, which can have orbital velocities $>$ $\sigma$, have been a long running concern in the measurements of $\sigma$ for dwarf spheroidal galaxies because those measurements  imply tremendous quantities of dark matter \citep{olszewski,minor} and for young stellar clusters \citep{gieles10b}. On the other hand, the issue has not raised much concern in the analysis of globular clusters because the inferred masses for the old clusters are in moderate agreement with expectations drawn from stellar models. Direct investigation of the binary fraction in a well-studied old cluster \citep{pryor} has confirmed a low binary fraction, but the fractions are found to be larger in young clusters \citep{port} and in low mass clusters \citep{milone}.
We need an estimate of the magnitude of the effect binaries might have on the measured $\sigma$'s of our clusters.

\cite{minor} did extensive modeling to address this issue for low mass dwarf galaxies. Except for
possible internal evolution in clusters, such as the destruction of binaries and mass segregation, the results of these calculations should be applicable here. They find that for a system with an intrinsic value of $\sigma = 4$ km s$^{-1}$, which is appropriate for our intermediate age clusters, binaries are likely to bias $\sigma$ upward by between 10 and 20\% depending on the binary fraction. Again, this modeling depends on the various properties of binary stars in addition to the fraction, but the fraction may actually be lower in clusters than adopted as typical elsewhere due to the increased disruption of binaries in clusters \citep{spitzer1}. Even though this level of bias is comparable to or slightly larger than the uncertainties we quote, it is small compared to the size of the effect that concerns us, which is at least a factor of $\sim  2$ larger. The effect should be further reduced in our measurements relative to their calculation because we place little weight on the wings of the LOSVD by fitting the line cores (this also helps mitigate against interloping stars). Finally, this bias is important in our discussion only if cluster $\sigma$'s are affected differentially as a function of age. 

Binaries are a challenging solution to the current problem because such a model requires 
careful coordination of the binary fraction as a function of age, given the smooth rise in $\Upsilon_*$ around ages of 1 Gyr that mimics evolution, and then the subsequent lack of observationally important binaries in older systems. 
Despite such an argument,  there is precedent for stellar evolution highlighting certain stellar populations, with different binary fractions, at different times \citep[see][]{gieles10b}. That work, however, focused on a population that was 10 Myr old, where it is easier to highlight short-lived populations. At ages of 1 Gyr a wider mass range of stars contribute significantly to the luminosity. We believe such a scenario unlikely, but cannot yet reject it categorically (see \S6 for how we expect to rule it out).

\subsection{Are Evolutionary Models the Problem?}
\label{sec:empirical}

The unexpected behavior of $\Upsilon_*$ with age could reflect a problem with our expectations. Previously unappreciated complications that affect the calculations of $\Upsilon_*$ continue to be highlighted \citep{fan}.
Due to the nature of the data in Figure \ref{fig:m2l}, such an approach must predict falling values of $\Upsilon_*$ with age for stellar populations with ages somewhere between 9.4 $<$ log(age [yrs]) $<$ 9.8. Such a 
prediction does not pose a problem for our sample of clusters, which does not probe these epochs, but it might for galaxies, which do. Studies of early-type galaxies, which are presumed to have had relatively simple star formation histories, have concluded that the evolution out to $z \sim 1$ is broadly consistent with the expectation of passive evolution for an intrinsically old population \citep[for examples see][]{vandokkum96,kelson,treu,treub,vandeven,vd07}. These results depend on observations from which the investigators infer that values of $\Upsilon_*$ are dropping with increasing redshift, or decreasing age, rather than rising.

\begin{figure}[htbp]
\begin{center}
\epsscale{0.9}
\plotone{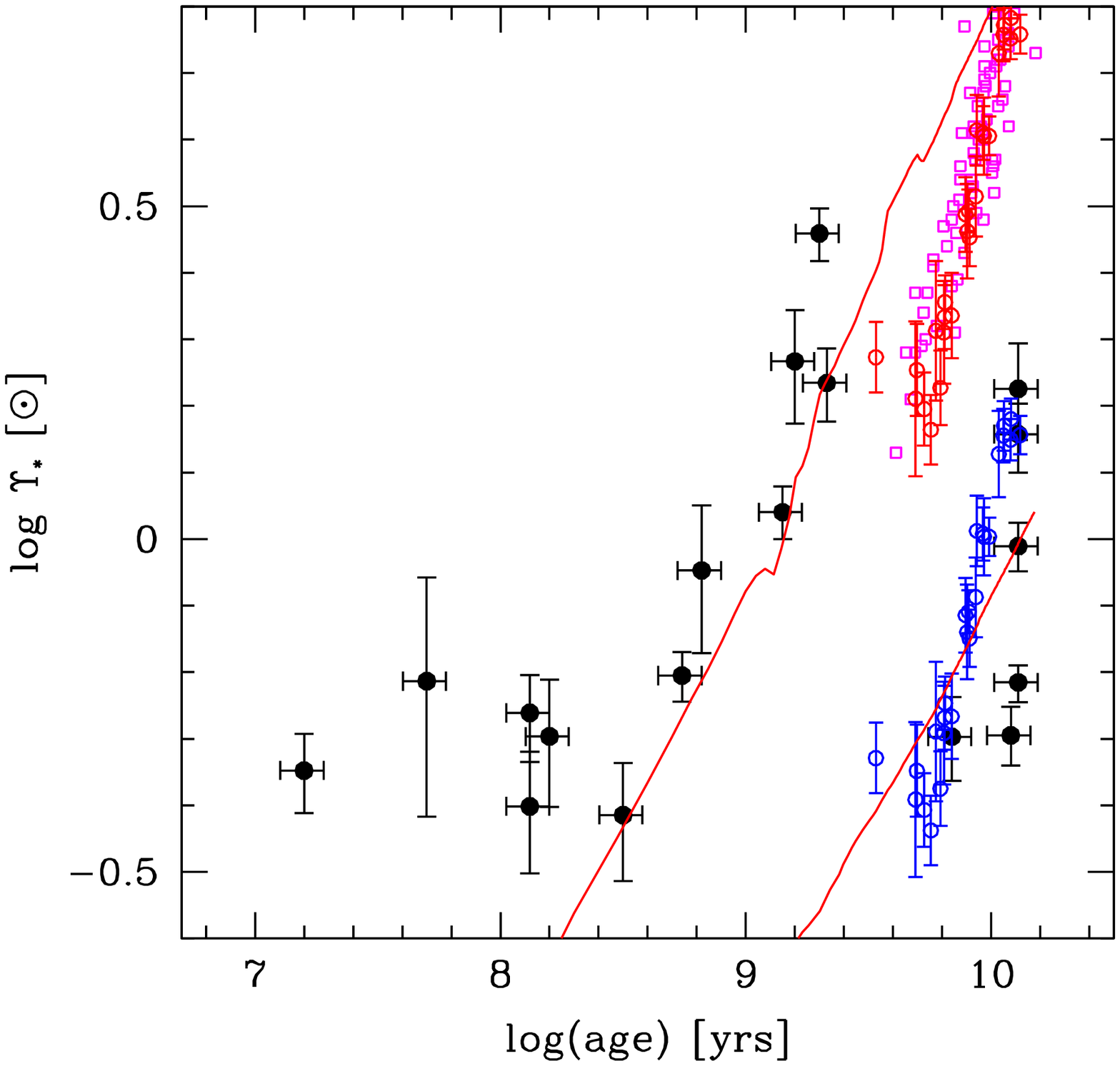}
\caption{The evolution of $\Upsilon_*$ as inferred from early type galaxies and comparison to models and our cluster data. Our cluster data are plotted in solid circles (black), the results obtained using the results from \cite{vandokkum} are plotted with open circles (red) when normalized to match the results from the sample of \cite{graves}, which are plotted with squares (pink). We correct the \cite{vandokkum} results to the $V$ band using a simple stellar population model, but no correction for dark matter (as such these are maximal $\Upsilon_*$). We also show the \cite{vandokkum} data assuming a dark matter fraction of 80\%, so as to qualitatively match to our stellar cluster data, in open circles (blue). The likely dark matter fraction is likely to be somewhere between 0 and 80\%. The upper solid line represents the values of $\Upsilon_h$ from a PEGASE model of a population with an instantaneous burst at $t=0$ and a Salpeter IMF, while the lower represents a model with a light-weighted Kroupa IMF.  In the log-log space used here, the model tracks can be approximated as straight lines of slope $-$0.77. }
\label{fig:imfvar}
\end{center}
\end{figure}

To determine whether galaxies can help discriminate against strong swings in $\Upsilon_*$ at intermediate ages,
we use measurements of the ages and structural parameters of two independent sets of early type galaxies. First, we use the study of Sloan Digital Sky Survey (SDSS) local galaxies presented by \cite{graves}. In this study, they binned early-type galaxies by their structural parameters, $\sigma$, $r_h$, and $I_h$, and examined deviations from the Fundamental Plane \citep{d87,dd87} relative to age and chemical abundance as measured from the stacked spectra of galaxies in that bin using the Balmer absorption lines and Lick indexes. These are luminosity weighted quantities and subject to the usual caveats involved in parameterizing complex populations by a single age and metallicity. Nevertheless, we use our Equation \ref{eq:fm} to calculate values of $\Upsilon_h$ and compare those values, as a function of age, to those of our stellar clusters. One important distinction is that unlike the clusters, these galaxies do contain dark matter and the exact proportion of dark matter within $r_h$ is unknown and likely to vary as a function of these structural parameters \citep[for some examples from the long history of this topic see][]{babul,graham,marinoni,vdb,cappellari,z06,wolf}. The results of our calculation are presented in Figure \ref{fig:imfvar}. Second, we use the data presented by \cite{vd07} who ascribed deviations from the Fundamental Plane for samples of galaxies in galaxy clusters as a function of redshift to evolution in $\Upsilon_*$. For these galaxies, we do not have a directly measured age estimate, so instead we assume 
that these galaxies formed at high redshift (here we take the formation time to be log(age) = 10.11, so that an elliptical galaxy at $z=0$ has the same age as our old stellar clusters) and that they have evolved passively thereafter. The age corresponding to a particular redshift is calculated using $H_0 =$ 70 km s$^{-1}$ Mpc$^{-1}$, $\Omega_m = 0.3$, and $\Omega_\Lambda = 0.7$. We correct their values of $\Upsilon_*$, which are in the rest $B$-band, to $V$ by using the calculated color evolution of a stellar population with an instantaneous burst of star formation at log(age) = 10.11 using PEGASE \citep{pegase} and adopting default values for the binary fraction (0.5), stellar wind prescriptions, and a \cite{miller} IMF that extends between 0.1 and 120 M$_\odot$. Because the measured shifts from the $z=0$ Fundamental Plane do not provide us with an absolute calibration of $\Upsilon_*$, the authors tabulate values for $\Delta \Upsilon_*$ as a function of $z$, there is some freedom in normalization. We normalize the values of $\Upsilon_h$ to match those calculated for the \cite{graves} galaxies, although an alternate normalization using the results of \cite{cappellari} places the \cite{vd07} slightly below the \cite{graves} galaxies.  In either case, the excellent agreement in the slope of the trend of $\Upsilon_h$ with age is important because the galaxy ages in these two samples are measured 
so differently.

These results (Figure \ref{fig:imfvar}, vertical axis changed to log($\Upsilon_*$) for a reason that will become apparent below) represent the upper bounds of $\Upsilon_*$ in early-type galaxies because a correction for the contribution of dark matter to $\Upsilon_h$ will drive $\Upsilon_*$ downward. The exact magnitude of the correction is uncertain because it depends on the very stellar evolutionary models that we are attempting to test. We can posit that dark matter is responsible for the entire offset between this population and our old stellar clusters, if we hypothesize that the stars in early-type galaxies are drawn from the same parent population as those in clusters. To arrive at this agreement would require an 80\% dark matter mass fraction, and we show the resulting comparison also in Figure \ref{fig:imfvar}. Perfect agreement is unlikely because the galaxies are less likely than the clusters to be described well as the result of instantaneous burst of star formation and the dark matter fraction will vary among galaxies. 

In the same Figure, we also include two model tracks for the evolution of simple stellar populations. First, we note that 
we have corrected the plotted values of $\Upsilon_*$ for the stellar clusters for the effect of internal dynamical evolution using the results of the \cite{anders} models.
We plot the results of PEGASE models using a Salpeter IMF, spanning from 0.1 to 120 M$_\odot$, with default stellar mass loss and binarity parameters, and metallicity matching the mean of the young clusters ($-0.4$). We have not renormalized these results in any way and they do an acceptable job of reproducing the trend seen in the younger clusters, for clusters of 8 $< $log(age [yrs]) $<$ 9.4.  As we discussed previously, we ignore discrepancies at younger ages due to the possibility that these clusters are not relaxed. The slight systematic underestimation of $\Upsilon_*$ could be addressed by extending the IMF to somewhat lower mass and perhaps adding some substellar mass, which is set to zero in the current models. We also plot results obtained using a Kroupa IMF, with the same parameters as the Salpeter model except that the metallicity is set to match that of the older clusters ($-1.8$) and here we are forced to renormalize the results by removing half of the stellar mass to reach agreement with the properties of the older clusters. 
We therefore refer to the function shown as a light-weighted Kroupa IMF. 

The comparisons between clusters, galaxies and models provide several interesting results.
First, we confirm that the early-type galaxies require that $\Upsilon_*$ decrease with lookback time rather than 
increase over the epoch where we lack cluster measurements. This result rules out 
models with strong swings in $\Upsilon_*$ at intermediate ages.
Second, the early-type galaxies lie between the extrapolation of the young cluster trend and the old clusters. This ``disagreement" is 
is easily addressed because the measured values of $\Upsilon_h$ are not equivalent to $\Upsilon_*$ due to dark matter's contribution to the mass.
Third, we are almost at the point with the galaxy sample where it overlaps in age with the intermediate age clusters. 
Such overlap would provide a valuable test of the models because
even though the two types of systems appear at the same age, they are inherently different in that for the clusters we are plotting stars formed $X$ years ago, while for the galaxies
we are plotting stellar populations that presumably formed 13 Gyr ago but that we are seeing as they were $X$ years after their formation.
Fourth, we find another reason for dismissing dramatic dynamical evolution of the old clusters as a solution to our problem because 
the values of $\Upsilon_h$ for the galaxies already lie below the extrapolation of the young clusters, and their $\Upsilon_*$ values will lie even lower due to the dark matter correction. 

\section{The Initial Mass Functions}
\label{sec:imf}

We conclude that the solution to the apparent discrepancy illustrated by Figures \ref{fig:m2lmodel} and \ref{fig:imfvar} is most directly addressed with the existence of at least two distinct stellar initial mass functions.

Although our data are limited by sample size, they can begin to inform a discussion regarding the cause of IMF variations. For example, one plausible candidate for the driver of those variations is metallicity, although in fact little variation in the IMF has been seen to date with metallicity and some theoretical understanding for that finding has emerged \citep{krumholz}. In our data, however, all of the old clusters are of low metallicity ($\langle$Fe/H$\rangle \sim - 2$) and NGC 416, which is the only cluster with log(age) $<$ 10 that lies with the old clusters is also of relatively low metallicity ($-1.44$). NGC 416 is a particularly interesting candidate in this context because it formed well after the initial episode of star formation in the SMC (being only 7 Gyr old). If we search the \cite{mclaughlin} data for other clusters that are of intermediate age 9 $<$ log(age [yrs]) $<$ 10 and low metallicity ($\langle$Fe/H$\rangle < -1$), we find only three additional such clusters (KRON 3, NGC 339, and NGC 361). Unfortunately, the mean surface brightness within r$_h$ of these clusters is at least 7 times fainter than that of NGC 416, which already required over 5000 sec of exposure time. An alternative test is to find old clusters (log(age [yrs]) $>$ 10) with high abundances ($\langle$Fe/H$\rangle > -0.5$). In this case we find 11 clusters (Liller 1, NGC 5927, NGC 6440, NGC 6528, NGC 6553, NGC 6624, Pal 8, Pal 10, Pal 11, Terzan 2, and Terzan 5). Of these, NGC 6440, 6528, 6553,  and Pal 11,  have the required data, other than $\sigma$, and high mean surface brightness, and are therefore good candidates for subsequent study.

\begin{figure}[htbp]
\begin{center}
\plotone{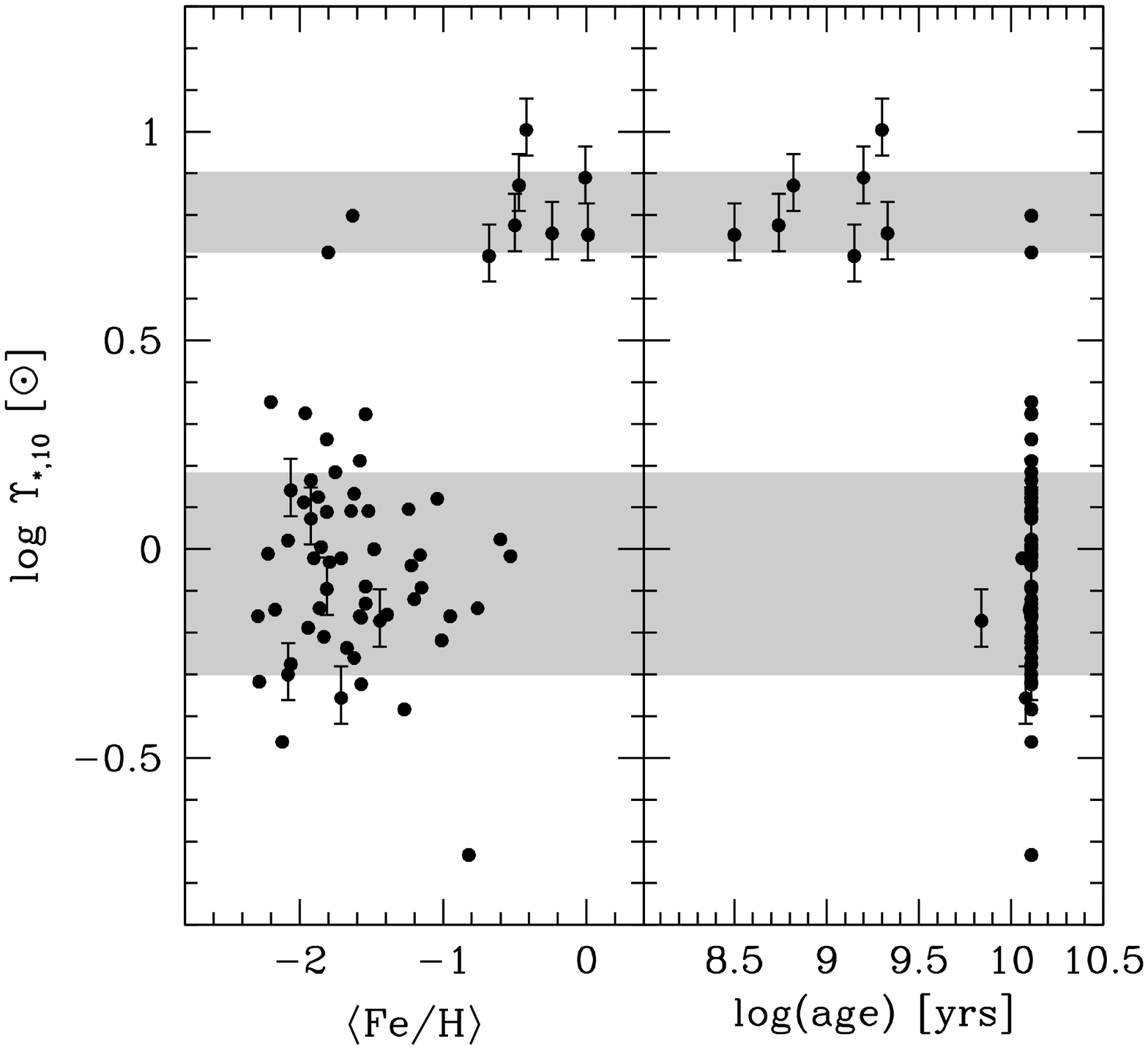}
\caption{Evidence for two initial mass functions. We plot  $\Upsilon_{*,10}$, the value of $\Upsilon_*$ at 10 Gyr for each cluster obtained as described in the text, versus iron abundance and age. The clusters included are those from our sample (with errorbars) and the \cite{mclaughlin} sample for log(age) $>$ 8.3. For the \cite{mclaughlin} clusters we adopt their calculated values of $\Upsilon_*$ from the dynamical models, except we reduce those estimates by a factor of 0.7 to match normalizations to our estimates (\S \ref{sec:comp}). For our clusters, we adopt the values of $\Upsilon_*$ that include the corrections for dynamical evolution obtained from the \cite{anders} models. The shaded regions represent the single cluster rms scatter of each population, centered on the mean value of $\Upsilon_{*,10}$ for each population.}
\label{fig:twoimf}
\end{center}
\end{figure}

To uncover any additional constraints on potential drivers of distinct IMFs, we combine our data and that compiled by \cite{mclaughlin}. In the log-log space of Figure \ref{fig:imfvar}, the model tracks are roughly linear with slope $-0.77$. As such,  it is straightforward to predict the value of $\Upsilon_*$ for each cluster at any age. We choose to calculate $\Upsilon_*$ at 10 Gyr, $\Upsilon_{*,10}$, for all of the clusters that are older than log (age [yrs]) $>$ 8.3 as a way to directly compare them. The differences among values of $\Upsilon_{*,10}$ will reflect differences in the IMF, if (1) the effects of dynamical evolution have been correctly accounted for, (2) stellar evolution models do follow the adopted linear behavior (in log space), and (3) clusters along any particular evolutionary track are of similar chemical abundance or chemical abundance does not grossly affect $\Upsilon_{*,10}$. We present the distribution of $\Upsilon_{*,10}$ versus age and chemical abundance in Figure \ref{fig:twoimf}.

The distribution of $\Upsilon_{*,10}$ is striking. There are two well-separated populations of clusters, highlighted by the grey bands that represent the mean and single cluster standard deviation for each of the two populations. The means of each population are statistically distinct at $\sim 10\sigma$. It is also evident that neither age nor $\langle$Fe/H$\rangle$ appear to be the sole arbiter of stellar population properties. Unfortunately, neither the clusters (NGC 2257 and 6535) at low $\langle$Fe/H$\rangle$ and high $\Upsilon_{*,10}$, which best overlap the older clusters, nor the low $\langle$Fe/H$\rangle$, low $\Upsilon_{*,10}$ clusters (NGC 104, 6362, 6388, and 6441, excluding NGC 6366 which has an anomalously low measurement of $\Upsilon_*$) that overlap the young population are among those for which we obtained measurements of the internal kinematics. As such, we cannot determine the confidence with which these are clusters that provide evidence for overlapping properties among these two populations. If their velocity dispersions are confirmed by our upcoming measurements, then these clusters would be key in unraveling the nature of the two populations. 

Direct measures of the mass function, obtained by counting stars, is evidently the most robust way to determine if there are indeed IMF variations among the clusters. Unfortunately, the Galactic clusters for which this has been done extremely well are all old clusters, and therefore lie predominantly in only one of our two populations. Nevertheless, these results can still shed some light on some of our results. First,
for NGC 6366, which is the outlier with extremely low $\Upsilon_{*,10}$ in Figure \ref{fig:twoimf}, the direct measurement of the stellar mass function from {\sl Hubble Space Telescope} imaging demonstrates that its mass function is indeed anomalous in that it has far fewer low mass stars than other clusters \citep{paust09}. This results at least qualitatively validates the low value shown in the Figure. \cite{paust09} attribute the low number of low mass stars, or correspondingly our low value of $\Upsilon_{*,10}$,  to strong tidal stripping because this cluster lies in the Galactic bulge. Second, excluding NGC 6366, the range of $\Upsilon_*$ among the group of low $\Upsilon_*$ clusters is roughly 0.5 dex or a factor of 3.
Using HST imaging of a set of Galactic clusters (which are all old and so presumably fall into this group), \cite{paust} find that the stellar mass function slopes, for $M < 0.8M_\odot$, vary from $-1.7$ to $-0.3$, where the Salpeter slope is $-2.35$ in this convention. This range of slopes corresponds to a factor of 3.6 change in the mass contained in stars between 0.1 and  0.8 M$_\odot$, and so to a somewhat smaller change in the overall mass. We conclude that much of the apparent scatter in 
$\Upsilon_{*,10}$ could be due to these slope differences, which are presumably due to different internal evolution. This finding is further confirmed with a sample of SMC clusters \citep{glatt}. Here the clusters span ages from 9.8$<$ log (age [yrs]) $< 10.2$, so again they unfortunately do not reach the clusters in our high $\Upsilon_*$ population, but have slopes ranging from $-$1.4 to $-$1.74, if one excludes two ambiguous determinations and a cluster that we discuss below. The two ambiguous results are for NGC 121 and NGC 416, two clusters that are also part of our sample, for which the slope is either around $-$1.4, which would be entirely consistent with the Galactic clusters, or $-$2.3, depending on whether the central region is excluded or replaced with higher resolution $HST$ HRC data. In one case the slope decreases when the inner data are included, in the other it increases, so the uncertainty here is large. A third cluster, NGC 339, with a steep mass function slope is potentially an interesting target because that one might lie in the high $\Upsilon_*$ population, although there is no measurement of $\sigma$ yet available. 

\section{Summary}

Using determinations of the velocity dispersions of 18 stellar clusters within the Local Group that span 7 $<$ log (age [yrs]) $<$ 10.2, we trace the evolution of the stellar mass-to-light ratio of simple stellar populations. We find that the observed behavior does not match simple theoretical expectations, and that the failure is such that global scalings of the masses, luminosities, or universal initial mass function cannot reconcile the two. The effect of the internal dynamical evolution of the clusters on $\Upsilon_*$ goes in the correct sense to alleviate the discrepancy, but fails to produce results that the sufficiently large. This is confirmed both by using models of the dynamical evolution \citep{anders} and direct measures of the stellar mass function in nearby clusters \citep{paust,leigh}. Contamination by binaries is also discussed, but seems problematic due to fine tuning problems. 

The data can be explained if there are two populations of clusters, drawn from different initial mass functions. Specifically, we find that
one IMF is primarily, but not exclusively, appropriate for older, metal poor clusters and the other for primarily, but not exclusively, for younger, metal rich clusters. The young (log(age [yrs])$<$9.5) clusters are well-described by a bottom-heavy IMF, such as a Salpeter IMF, while the older clusters are better described by a top-heavy IMF, such as a light-weighted Kroupa IMF, although neither of these specific forms is a unique solution.  Of course, the sample is currently small and there are only four clusters in the key age range. We are in the process of enlarging the sample, although there are not many intermediate age clusters available for such study with the current technology.
There are also a variety of effects, including the dynamical evolution described above, whose magnitude is estimated using models; therefore, significant departures from those models can also be used to explain the observations.

Claims for variations in the initial mass function are usually treated with significant skepticism. This is not because we have a basic understanding of the initial mass function that predicts universality, but is rather due to a fondness for Occam's razor. In fact, certain previous claims of variations in the IMF are in the same sense as the effect seen here. For example, \cite{kroupa01} found evidence for greater numbers of low mass stars in younger Galactic populations and \cite{dave} inferred top heavy star formation at earlier times on the basis of galaxy evolution models. Whether these findings are 
quantitatively consistent with what we present is unclear, but certainly worth exploring to determine if a consistent picture can emerge.

 It is certainly the case that allowing for the possibility of varying IMFs tremendously complicates what one can constrain in extragalactic astronomy. Even if there are only two IMFs, the variation between those as a function of age, environment, star formation rate, or whatever else drives the dichotomy provides numerous new degrees of freedom to any simulation or model. Of course, some have already appealed to such effects to alleviate conflicts between models and data, but multiple IMFs not only enables one to address what were previously frustrating conflicts, but requires one to reassess what were previously apparent successes. Even worse, without knowing the underlying physical cause for the two classes of IMFs claimed to be responsible for the observations presented here, we cannot claim that there are only two variants. Unfortunately, this is now an added complication that we must face or at least include as a source of systematic error in any study involving galaxy luminosities, color, or stellar mass determinations.

\begin{acknowledgments}

DZ acknowledges financial support from 
NSF grant AST-0907771 and NASA ADAP NNX12AE27G and thanks the Max Planck Institute for Astronomy for its hospitality during the completion of this study. R. C. acknowledges support from NSF through CAREER award 0847467.

\end{acknowledgments}

\end{document}